\documentclass[11pt,a4paper]{article}

\usepackage{amsmath}
\usepackage{amsfonts}
\usepackage{color}
\usepackage{cite}
\usepackage{graphicx}
\usepackage{float}

\UseRawInputEncoding

\title{\textbf{Asymmetric scattering between kinks and wobblers}}

\author{A. Alonso-Izquierdo$^{(a)}$, L. M. Nieto$^{(b)}$ and J. Queiroga-Nunes$^{(a)}$
\\ $^{(a)}$ Departamento de Matematica Aplicada, Universidad de Salamanca, \\ Casas del Parque 2, 37008 - Salamanca, Spain \\
and IUFFyM, Universidad de Salamanca, \\ Plaza de la Merced 1, 37008 - Salamanca, Spain \\
$^{(b)}$ Departamento de F\'{\i}sica Te\'orica, At\'omica y \'Optica, and IMUVA, \\
Universidad de Valladolid, 47011, Valladolid, Spain}

\date{}

\begin{document}

\maketitle

\begin{abstract}
The asymmetric scattering between wobblers and kinks in the standard $\phi^4$ model is numerically investigated in two different scenarios. First, the collision between wobblers with opposite phases is analyzed. Here, a destructive interference between the shape modes of the colliding wobblers takes place at the impact time. The second scenario involves the scattering between a wobbler and an unexcited kink. In this case the energy transfer from the wobbler to the kink can be examined. The dependence of the final velocities and wobbling amplitudes of the scattered wobblers on the collision velocity and on the initial wobbling amplitude is discussed. Both situations lead to very different fractal structures in the velocity diagrams.
\end{abstract}

\section{Introduction}

The scattering between kinks has become a very popular research topic in recent decades because of its astonishing properties \cite{Manton2004, Shnir2018, Kevrekidis2019}. The study of the collisions between kinks and antikinks in the $\phi^4$ model was initially addressed in the seminal references \cite{Sugiyama1979, Campbell1983, Anninos1991}. As it is well known, only two different scattering channels arise: \textit{bion formation} (where kink and antikink collide and bounce back over and over emitting radiation in every impact) and \textit{kink reflection} (where kink and antikink collide and bounce back a finite number of times before moving away). These two channels are predominant, respectively, for low and large values of the initial collision velocity. In these studies emerges the fascinating property that the two previously mentioned channels are infinitely interlaced in the transition of these regimes, giving rise to a fractal structure embedded in the final versus initial velocity diagram. The kink reflection windows included in this region involve scattering processes where kink and antikink collide and bounce back a finite number of times before definitely escaping away. This kink dynamics could have important consequences on physical applications where the presence of these topological defects allows the understanding of certain non-linear phenomena. Kinks (and topological defects in general) have been employed in a wide variety of physical disciplines, such as Condensed Matter \cite{Eschenfelder1981,Jona1993,Strukov}, Cosmology \cite{Vilenkin1994,Vachaspati2006}, Optics \cite{Mollenauer2006,Schneider2004,Agrawall1995}, molecular systems \cite{Davydov1985,Bazeia1999}, Biochemistry \cite{Yakushevich2004}, etc.

The appearance of a fractal structure in the velocity diagram describing the kink scattering for the $\phi^4$ model is based on the existence of an internal vibrational mode (the shape mode) associated to the kink solutions. The presence of this massive mode together with the zero mode triggers the \textit{resonant energy transfer mechanism}, which allows the redistribution of the energy between the kinetic and vibrational energy pools when the kinks collide. In a usual scattering event the kink and the antikink approach each other and collide. A certain amount of kinetic energy is transferred to the shape mode, such that kink and antikink become wobblers (kinks whose shape modes are excited), which try to escape from each other. If the kinetic energy of each wobbler is not large enough both of them end up approaching and colliding again. This process can continue indefinitely or finish after a finite number of collisions. In this last case, enough vibrational energy is returned back to the zero mode as kinetic energy, which allows the wobblers to escape. This mechanism and other related phenomena have been thoroughly analyzed in a large variety of models \cite{Shiefman1979, Peyrard1983, Goodman2005, Gani1999, Malomed1989, Gani2018, Gani2019,Simas2016,Gomes2018,Bazeia2017b,Bazeia2017a, Bazeia2019, Adam2019, Romanczukiewicz2018, Adam2020, Mohammadi2020, Yan2020,Romanczukiewicz2017, Weigel2014, Gani2014, Bazeia2018b, Lima2019, Marjaheh2017, Belendryasova2019, Zhong2020, Bazeia2020c, Christov2019, Christov2019b, Christov2020,Halavanau2012, Romanczukiewicz2008, Alonso2018, Alonso2018b,Alonso2017, Alonso2019, Alonso2020, Alonso2021, Ferreira2019, Goodman2002, Goodman2004,Malomed1985,Malomed1992, Saadatmand2015,Saadatmand2018, Manton1997,Adam2018,Adam2019b,Adam2020b,Dorey2011,Dorey2018,Mohammadi2021b,Campos2020,Blanco-Pillado2021}, revealing the enormous complexity of these events and the difficulty in explaining this phenomenon analytically. The \textit{collective coordinate approach} has been used to accomplish this task for decades, reducing the field theory to a finite dimensional mechanical system, where the separation between the kinks and the wobbling amplitudes associated to the shape modes are promoted to dynamical variables. This method has been progressively improved, see for example \cite{Sugiyama1979,Takyi2016, Kevrekidis2019,Pereira2020} and references therein, and recently, a reliable description of the kink scattering in the $\phi^4$ model has been achieved in the reflection-symmetric case \cite{Manton2021}, by introducing in this scheme the removal of a coordinate singularity in the moduli space and choosing the appropriate initial conditions.

As previously mentioned, after the first collision the initially unexcited kink and antikink become wobblers, so in a $n$-bounce scattering process the subsequent $n-1$ collisions can be understood as scattering processes between two wobblers. This observation justifies an intrinsic interest on the collision between these objects. The evolution of a single wobbler has been studied by employing perturbation expansion schemes by different authors, see \cite{Barashenkov2009,Barashenkov2018,Segur1983} and references therein. The scattering between wobblers in the $\phi^4$ model has been discussed in \cite{Alonso2021b} for a space reflection symmetric scenario. This situation is relevant in the original kink scattering problem where the mirror symmetry is preserved. The goal of these investigations is to bring insight into the resonant energy transfer mechanism by means of numerical analysis of the scattering solutions derived from the corresponding the Klein-Gordon partial differential equations. In this context it is worthwhile mentioning that the scattering of wobblers in the double sine-Gordon model has been studied by Campos and Mohammadi \cite{Campos2021}.

In this paper we shall continue with this line of research by investigating the asymmetric scattering between wobblers in two different scenarios, which are considered representative of this context. The scattering processes addressed in previous works involve wobblers which evolve with the same phase. This implies that a constructive interference between the shape modes associated to each wobbler takes place at the collision. In this work we propose the analysis of the scattering between wobblers with opposite phases, such that now a destructive interference between the vibrational modes occurs at the impact. The second scenario is described by the collision between a wobbler and an unexcited kink. This allows us to monitor the transfer of the vibrational energy from the wobbler to the kink. We will show that the fractal structures ruled by the resonance phenomenon in these two cases display very different patterns.

The organization of this paper is as follows: in Section \ref{sec:2} the theoretical background of the $\phi^4$ model together with the analytical description of kinks and wobblers is introduced. The kink-antikink scattering is also discussed, which allows us to describe the numerical setting employed to study the problem. Section~\ref{sec:3} is dedicated to study the scattering between wobblers with opposite phase, whereas the collision between a wobbler and an unexcited kink is addressed in Section \ref{sec:4}. Finally, some conclusions are drawn in Section~\ref{sec:5}.

\section{The $\phi^4$ model: kinks and wobblers}

\label{sec:2}

The dynamics of the $\phi^4$ model in (1+1) dimensions is governed by the action
\begin{equation}\label{action}
S=\int d^2 x \,\, {\cal{L}}(\partial_{\mu}\phi, \phi) \hspace{0.5cm},
\end{equation}
where the Lagrangian density ${\cal{L}}(\partial_{\mu}\phi, \phi)$ is of the form
\begin{equation}\label{lagrangiandensity}
{\cal{L}}(\partial_{\mu}\phi, \phi) = \frac{1}{2} \,\partial_\mu \phi \, \partial^\mu \phi - V(\phi) \hspace{0.5cm} \mbox{with} \hspace{0.5cm} V(\phi) = \frac{1}{2} (\phi^2 -1)^2 \hspace{0.5cm}.
\end{equation}
The use of dimensionless field and coordinates, as well as Einstein summation convention, are assumed in expressions (\ref{action}) and (\ref{lagrangiandensity}). Here, the Minkowski metric $g_{\mu\nu}$ has been set as $g_{00}=-g_{11}= 1$ and $g_{12}=g_{21}=0$. Therefore, the non-linear Klein-Gordon partial differential equation
\begin{equation}
\frac{\partial^2 \phi}{\partial t^2} - \frac{\partial^2 \phi}{\partial x^2} = -2\phi(\phi^2-1) 
\label{pde}
\end{equation}
characterizes the time-dependent solutions of this model. The energy-momentum conservation laws imply that the total energy and momentum
\begin{equation}
E[\phi] = \int dx \Big[ \frac{1}{2} \Big( \frac{\partial \phi}{\partial t} \Big)^2 + \frac{1}{2} \Big( \frac{\partial \phi}{\partial x} \Big)^2  + V(\phi) \Big] \hspace{0.5cm}, \hspace{0.5cm}
P[\phi] = - \int dx\, \frac{\partial \phi}{\partial t} \, \frac{\partial \phi}{\partial x}  \hspace{0.5cm}, \label{invariants}
\end{equation}
are system invariants. The kinks/antikinks ($+/-$)
\begin{equation}
	\phi_{\rm K}^{(\pm)}(t,x;x_0,v_0) = \pm \tanh \left[\frac{x-x_0-v_0 t}{\sqrt{1-v_0^2}}\right]  \label{travelingkink}
\end{equation}
are travelling solutions of (\ref{pde}), whose energy density is localized around the kink center $x_C=x_0+v_0 t$ (the value where the field profile vanishes). The parameter $v_0$ can be interpreted as the kink velocity. As it is well known, the solutions (\ref{travelingkink}) are topological defects because they asymptotically connect the two elements of the set of vacua ${\cal M}=\{-1,1\}$. These solutions have a normal mode of vibration. When this mode is excited the size of these solutions (called \textit{wobbling kinks} or \textit{wobblers}) periodically oscillates with frequency $\omega=\sqrt{3}$. This fact has been numerically checked and has been analytically proved in the linear regime. The spectral problem
\[
{\cal H} \psi_{\omega^2}(x) = \omega^2 \psi_{\omega^2}(x)
\] 
of the second order small fluctuation operator associated with the static kink/antikink,
\begin{equation}
{\cal H} = - \frac{d}{dx^2} + 4-6\,{\rm sech}^2 (x-x_0), \label{hessian}
\end{equation}
involves the shape mode
\[
\psi_{\omega^2=3}(x;x_0)=   \, {\rm sinh}\, (x-x_0) \, {\rm sech}^2 (x-x_0)
\]	
with eigenvalue $\omega^2=3$. The discrete spectrum of the operator (\ref{hessian}) is completed with the presence of a zero mode
\[
	\psi_{\omega^2=0}(x;x_0)=   \, {\rm sech}^2 (x-x_0) = \left. \frac{\partial \phi_K^{(+)}}{\partial x}\right|_{t=0,v_0=0} \hspace{0.5cm},
\]
whereas the continuous spectrum emerges on the threshold value $\omega^2=4$. 

As a result of this linear analysis, the expression
\begin{equation}
\phi_{\rm W}^{(\pm)}(t,x;x_0,v_0,\omega,a,\delta)  =  \pm \tanh \left[ \frac{x-x_0 - v_0 t}{\sqrt{1-v_0^2}} \right] +  a  \sin(\omega t+\delta){\rm{sech}} \left[ \frac{x-x_0 - v_0 t}{\sqrt{1-v_0^2}} \right]  \tanh\left[ \frac{x-x_0 - v_0 t}{\sqrt{1-v_0^2}} \right]  \label{wobbler}
\end{equation}
can be considered a good approximation of a traveling wobbler in the linear regime $a\ll 1$. Note that $\phi_{\rm W}^{(-)}(t,x)$ describes a \textit{wobbling antikink} (or \textit{antiwobbler}).

The maximum deviation of the wobbler (\ref{wobbler}) from the kink (\ref{travelingkink}) takes place at the points
\begin{equation}
x_M^{(\pm)} = x_C \pm \sqrt{1-v_0^2} \,\,{\rm arccosh}\, \sqrt{2} \hspace{0.5cm} , \label{points}
\end{equation}
where the relation
\[
\left| \phi_{\rm W} (x_M^{(\pm)}) - \phi_{\rm K} (x_M^{(\pm)})\right| = \frac{1}{2} \, |a|
\]
holds. An optimized strategy to measure the wobbling amplitude of a traveling wobbler in a numerical scheme is to monitor the profile of these solutions at the points (\ref{points}). By using fourth order perturbation theory in the expansion parameter $a$, it has been proved that $a$ depends on time, $a=a(t)$, and decays following the expression
\begin{equation}
|a(t)|^2 = \frac{|a(0)|^2}{1+\omega \,\xi_I\, |a(0)|^2 t}, \label{amplitude}
\end{equation}
where $\xi_I$ is a constant. However, when the initial wobbling amplitude $a(0)$ is small, the decay is very slow and becomes appreciable only after a long time $t\sim |a(0)|^{-2}$ \cite{Barashenkov2009,Barashenkov2018}.

\begin{figure}[htb]
	\centerline{\includegraphics[height=3.5cm]{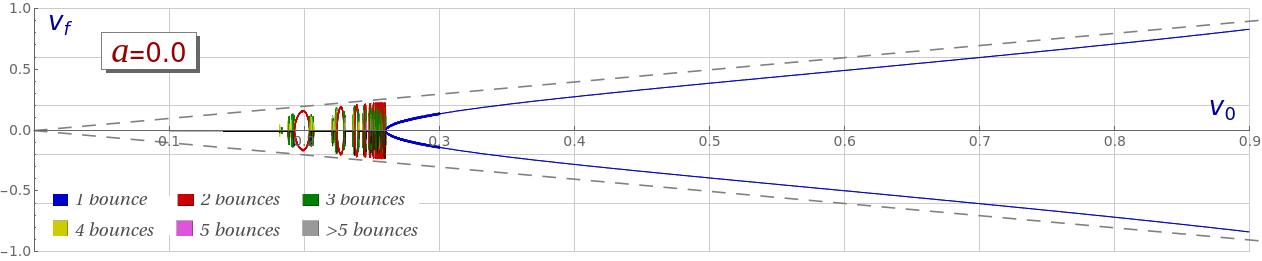}}
	\caption{\small Final versus initial velocity diagram for the kink-antikink scattering. The final velocity of the bion is assumed to be zero in this context. The color code is used to specify the number of bounces suffered by the kinks before escaping.} \label{fig:VelDiaAmp000}
\end{figure}

The scattering between a kink and an antikink has been thoroughly analyzed in the physical and mathematical literature during the last decades. In this case, a kink and antikink which are well separated are pushed together with initial collision velocity $v_0$. Taking into account the spatial reflection symmetry of the system the kink can be located at the left of the antikink or vice versa. For very small values of the time $t$ (with respect to the impact time), the previous scenario is characterized by the concatenation
\begin{equation}
\Phi_{KK}(t,x;x_0,v_0) =  \phi_K^{(\pm)}(t,x;x_0,v_0) \cup \phi_K^{(\mp)}(t,x;-x_0,-v_0) \label{configuration01}
\end{equation}
for $x_0\gg 0$, where we have introduced the notation
\begin{equation}
\phi_K^{(\pm)}(t,x;x_0,v_0) \cup \phi_K^{(\mp)}(t,x;-x_0,-v_0)\equiv \left\{ \begin{array}{ll} \phi_K^{(\pm)}(t,x;x_0,v_0)  & \mbox{if } x\leq 0, \\[0.2cm] \phi_K^{(\mp)}(t,x;-x_0,-v_0) & \mbox{if } x> 0  . \end{array} \right.  \label{concadef}
\end{equation}
The initial separation distance between the kink and the antikink is equal to $2x_0$. The configuration (\ref{configuration01}) defines the initial conditions of the scattering problem. As it is well known, there exist two different scattering channels in this case: (1) \textit{bion formation}, where kink and antikink end up colliding and bouncing back over and over, and (2) \textit{kink reflection}, where kink and antikink collide,  bounce, and finally recede with respective final velocities $v_{f,L}$ and $v_{f,R}$ in the opposite direction in which they were initially traveling. These scattering regimes are predominant, respectively, for low and high values of the initial velocity $v_0$. In Figure \ref{fig:VelDiaAmp000} the two previously mentioned final velocities $v_{f,L}$ and $v_{f,R}$ are plotted as a function of the initial collision velocity $v_0$.

From the spatial reflection symmetry exhibited by the initial configuration  (\ref{configuration01}) it is clear that $v_{f,L}=-v_{f,R}$ and that the velocity of a bion must be zero. Therefore, the velocity diagram in Figure \ref{fig:VelDiaAmp000} is symmetric with respect to the $v_0$-axis. In the next sections we shall address asymmetric scattering events where this symmetry is lost and $|v_{f,L}|\neq |v_{f,R}|$ in general. The fascinating property found in this scattering problem is that the transition between the two previously mentioned regimes is ruled by a fractal structure where the bion formation and the kink reflection regimes are infinitely interlaced. The kink reflection windows included in this initial velocity interval involve scattering processes where kink and antikink collide and bounce back a finite number of times exchanging energy between the zero and shape modes before definitely moving away. These processes involve the so called \textit{resonant energy transfer mechanism}.

For the previously mentioned $n$-bounce processes (with $n\geq 2$) it is clear that after the first impact the subsequent collisions correspond to scattering processes between wobblers because, in general, the collision between kinks causes the excitation of their shape modes. Taking into account the spatial reflection symmetry of the problem, the wobbling amplitudes and phases of the colliding wobblers are equal. Therefore, these events are characterized by an initial configuration of the form 
\begin{equation}
\Phi_{WW}(t,x;x_0,v_0,\omega,a,\delta)  = 
\phi_W^{(\pm)}(t,x; x_0,v_0,\omega,a,\delta) \cup 
\phi_W^{(\mp)}(t,x; -x_0,-v_0,\omega,a,\delta) \,. \label{configuration02}
\end{equation}
This scattering problem has been numerically studied in \cite{Alonso2021b}. By mirror symmetry, it can be assumed that the phases of the shape modes of the traveling wobblers are also the same at the impact time, so a constructive interference takes places in the collision. As a consequence, it was found that the fractal pattern enlarges and becomes more complex as the value of the initial wobbling amplitude $a$ increases. Another interesting property in this context is the emergence of isolated 1-bounce windows, which are not present in the original kink-antikink scattering. It is clear that the scattering between wobblers characterized by the initial configuration (\ref{configuration02}) is extremely relevant to study the resonant energy transfer mechanism in this problem. However, because of the spatial reflection symmetry of this type of processes, wobblers transfer the same amount of energy to each other at the collision, that is, the scattered wobblers travel away with the same final speeds and wobbling amplitudes. In this work we are interested in analyzing more general scattering events where the energy transfer mechanism becomes asymmetric with respect to the traveling wobblers.

The first type of processes which could involve novel properties in this framework is the collision between two wobblers with opposite phase. This scenario can be characterized by the initial configuration 
\begin{equation}
\Phi_{W \widetilde{W}}(t,x;x_0,v_0,\omega,a,\delta)  = 
\phi_W^{(\pm)}(t,x; x_0,v_0,\omega,a,\delta) \cup 
\phi_W^{(\mp)}(t,x; -x_0,-v_0,\omega,a,\pi+\delta) \, . \label{configuration04}
\end{equation}
We have employed the notation $W\widetilde{W}$ as subscript of $\Phi$ in (\ref{configuration04}) to simply emphasize that the wobblers have different initial phases and to distinguish this configuration from (\ref{configuration02}). In this case it is assumed that the wobblers evolve preserving a phase difference of $\pi$, giving place to a destructive interference in the excitation of the shape modes of each wobbler when they collide. It is expected that the final versus initial velocity diagrams associated to these scattering events will be affected by this fact and that they will be very different from those found in the constructive interference scenario (\ref{configuration02}), analyzed in \cite{Alonso2021b}.

Another important situation which deserves attention is the scattering between a wobbler and a kink. These asymmetric events can be characterized by the initial configuration
\begin{equation}
\Phi_{WK}(t,x;x_0,v_0,\omega,a,\delta)  =
\phi_W^{(\pm)}(t,x; x_0,v_0,\omega,a,\delta) \cup 
\phi_K^{(\mp)}(t,x; -x_0,-v_0) \, , \label{configuration03}
\end{equation}
where without loss of generality the non-excited antikink/kink $\phi_K^{(\mp)}(t,x; -x_0,-v_0)$ has been placed to the right of the wobbler/antiwobbler. This situation allows to analyze how the vibrational energy is transferred to the non-excited kink in a better way than in the previous contexts.

In order to study the scattering between kinks and wobblers in the two previously described scenarios, in the present work we shall employ numerical approaches based on the discretization of the partial differential equation (\ref{pde}) with different initial conditions determined by the configurations (\ref{configuration04}) and (\ref{configuration03}). The particular numerical scheme used here is a fourth-order explicit finite difference algorithm implemented with fourth-order Mur boundary conditions, which has been designed to address non-linear Klein-Gordon equations, see the Appendix in \cite{Alonso2021}. The linear plane waves are absorbed at the
boundaries in this numerical scheme avoiding that radiation is reflected in the simulation contours. To rule out the presence of spurious phenomena attributable to the use of a particular numerical algorithm, a second numerical procedure is used to validate the results. This double checking has been carried out by means of an energy conservative second-order finite difference algorithm with Mur boundary conditions.

As previously mentioned, the initial settings for our scattering simulations are described by single solutions (kinks or wobblers) which are initially well separated, and are pushed together with initial collision velocity $v_0$. This situation is characterized by the concatenation (\ref{configuration04}) for the scattering between wobblers with opposite phase and by (\ref{configuration03}) for the scattering between a wobbler and a kink, both of them with $x_0\gg 0$. These configurations verify the partial differential equation (\ref{pde}) in a very approximate way for very small values of the time when $x_0\gg 0$ and $a\ll 1$. Therefore, $\Phi(t=0)$ and $\frac{\partial \Phi}{\partial t}(t=0)$ provide the initial conditions of our scattering problem. 

In particular, our numerical simulations have been carried out in a spatial interval $x\in [-100,100]$ where the centers of the single solutions are initially separated by a distance $d=2x_0=30$. Simulations have been performed for $v_0\in [0.04,0.9]$ with initial velocity step $\Delta v_0=0.001$, which is decreased to $\Delta v_0=0.00001$ in the resonance interval.

At this point it is worthwhile mentioning that the expression (\ref{wobbler}) is only an approximation of the exact wobbler solution. When this expression is employed as initial condition in the Klein-Gordon equation (\ref{pde}) a small amount of radiation is emitted for a very small period of time. In this time interval the approximate solution (\ref{wobbler}) decays to the exact wobbler. When considering a traveling wobbler, this radiation emission can cause a very small change in its velocity. This effect takes place when $\delta\neq 0,\pi$ in the expression (\ref{wobbler}) and it is maximized for $\delta= \pm \pi/2$. In order to avoid this effect we shall implement initial conditions by setting $\delta=0$ in the configurations (\ref{configuration04}) and (\ref{configuration03}). By taking this restriction we guarantee that the traveling wobbler involved in (\ref{configuration03}) continues to move with velocity $v_0$ after the initial radiation emission. As mentioned above, this effect is very small and unnoticeable in the final versus initial velocity diagrams. However, we shall analyze the velocity difference of the resulting wobblers and in this context it is better to avoid this influence. On the other hand, for the values $\delta=0$ or $\delta=\pi$ the decay of the approximation (\ref{wobbler}) to the real wobbler induces a very small variation in its wobbling amplitude. This effect also is very small and does not affect the global properties of the scattering processes discussed in this paper. An alternative scheme to implement initial configurations (\ref{configuration03}) with non-vanishing initial phases is to find an approximately equivalent configuration with vanishing phase. This can be obtained, for example, taking into account that 
\[
\phi_W^{(\pm)}(t,x; x_0,v_0,\omega,a,\delta) = \phi_W^{(\pm)}(t-\textstyle\frac{\delta}{\omega},x; x_0-\frac{\delta v_0}{\omega},v_0,\omega,a,0)\hspace{0.3cm} .
\]

\section{Scattering between wobblers with opposite phases}
\label{sec:3}

\begin{figure}[h]
	\centerline{\includegraphics[height=3.5cm]{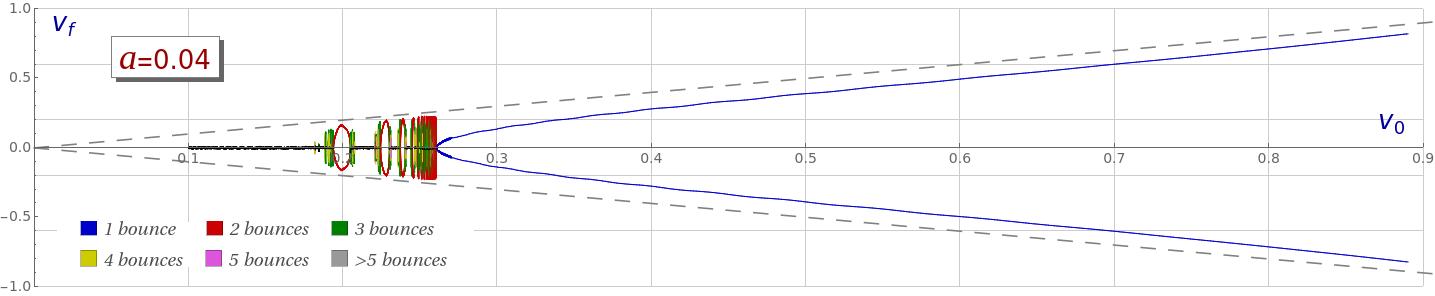}}
	\medskip
	\centerline{\includegraphics[height=3.5cm]{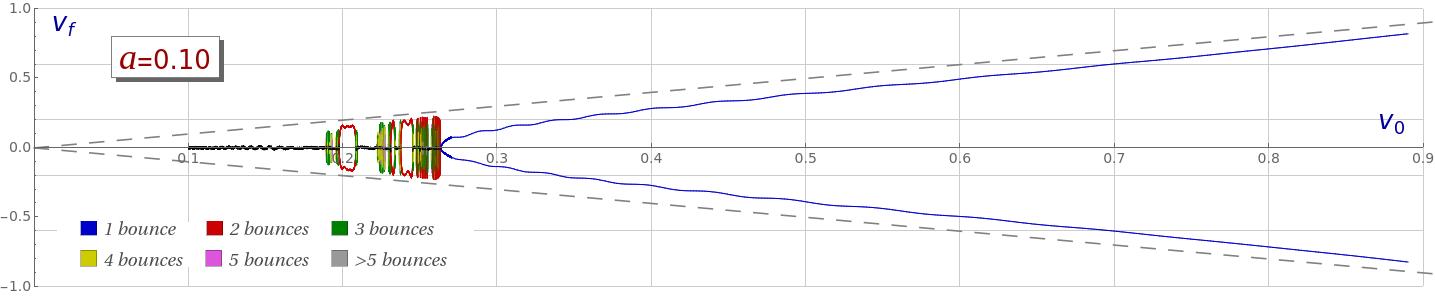}}
		\medskip

	\centerline{\includegraphics[height=3.5cm]{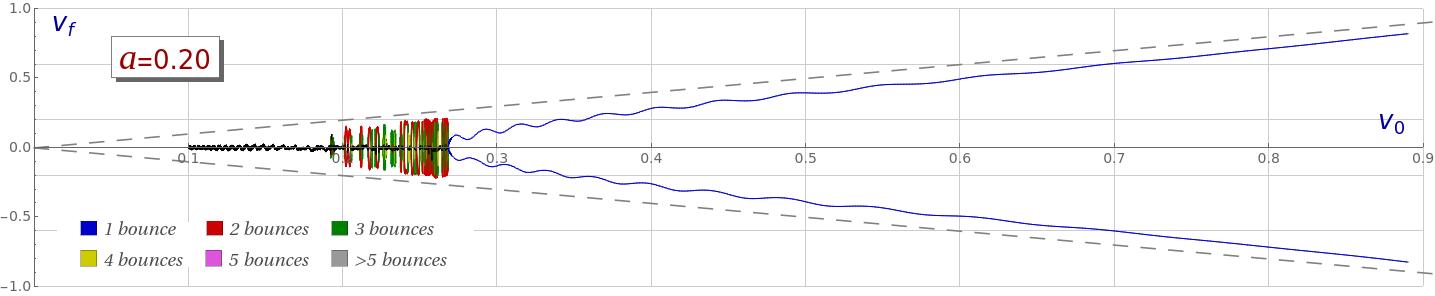}}

	\caption{\small Final versus initial velocity diagram for the scattering between two wobblers with opposite phase for the values of the wobbling amplitude $a=0.04$, $a=0.1$ and $a=0.2$. The color code is used to specify the number of bounces suffered by the kinks before escaping. The black curves determine the final velocity of the bion formed for low initial velocities.} \label{fig:VelDiaContrafaseAmp020}
\end{figure}

In this section we shall analyze the asymmetric scattering between two wobblers whose shape modes have the same amplitude but they have opposite phases with respect to our inertial system, which is  located at the center of mass. In this context, a wobbler and an anti-wobbler approach each other with initial velocity $v_0$ and $-v_0$, respectively. They evolve preserving the  phase difference of $\pi$ and collide giving place to a destructive interference between the shape modes of the involved wobblers. Figure~\ref{fig:VelDiaContrafaseAmp020} shows the final versus initial velocity diagrams for three representative values of the initial wobbling amplitude, $a=0.04$, $a=0.1$, and $a=0.2$.

\begin{figure}[htb]
	\centerline{\includegraphics[height=2.5cm]{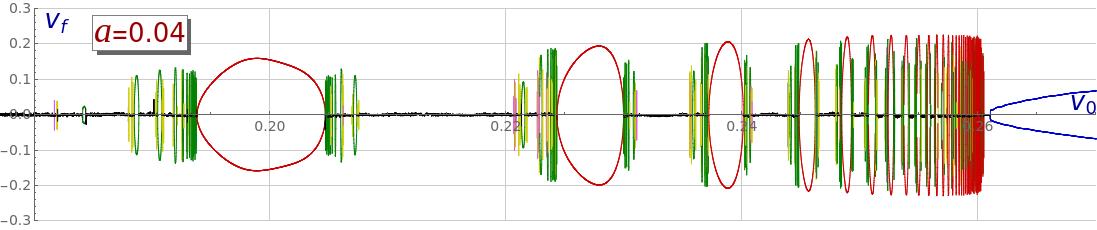}}
	\medskip
	\centerline{\includegraphics[height=2.5cm]{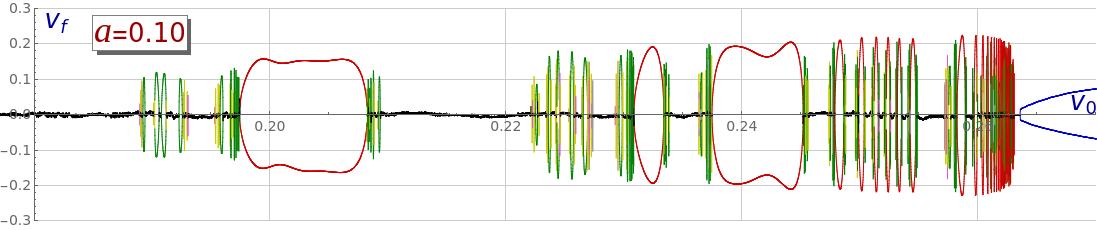}}
	\medskip
	\centerline{\includegraphics[height=2.5cm]{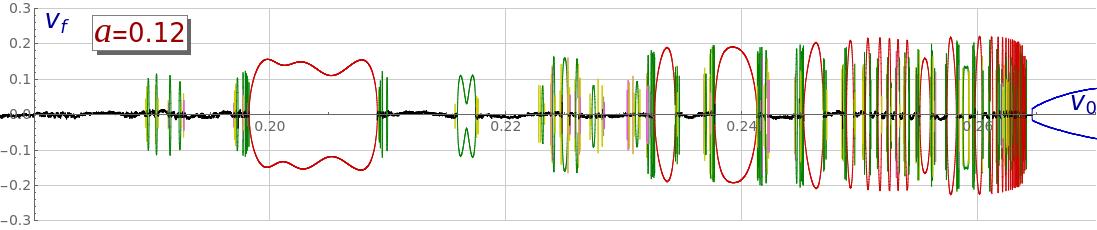}}
	\medskip
	\centerline{\includegraphics[height=2.5cm]{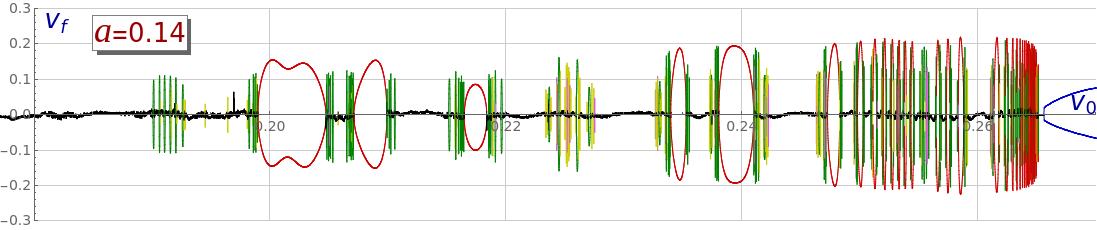}}
	\medskip
	\centerline{\includegraphics[height=2.5cm]{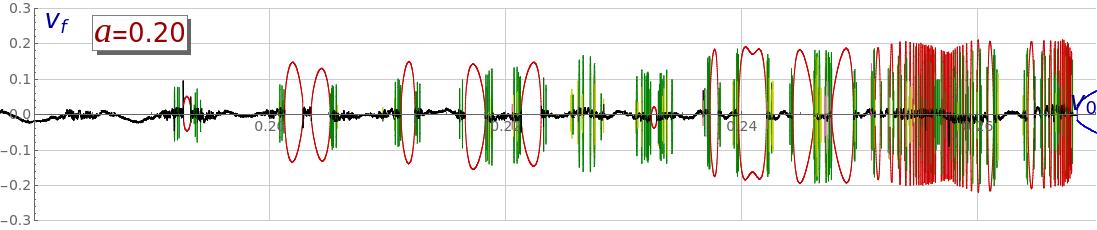}}
	\caption{\small Evolution of the fractal pattern found in the velocity diagrams associated to the scattering between wobblers with opposite phases as a function of the initial wobbling amplitude $a$.} \label{fig:VelDiaContrafaseFractal}
\end{figure}

Unsurprisingly, the only scattering channels to emerge in this new scenario are still bion formation and kink reflection.  As before, the former is predominant for small values of the initial velocity $v_0$, while the latter is found for large values. However, these velocity diagrams display some important differences regarding the scattering of wobblers addressed in \cite{Alonso2021b}, where the corresponding shape modes have the same phase and a constructive interference occurs in the collision.  In this new context, the destructive interference avoids the emergence of isolated 1-bounce windows (at least for non-extreme values of $a$), as can be observed in Figures~\ref{fig:VelDiaContrafaseAmp020} and \ref{fig:VelDiaContrafaseFractal}. The suppression of this mechanism implies that the fractal structure width does not grow. 
 
 In Figure \ref{fig:VelDiaContrafaseFractal} the evolution of the fractal pattern can be visualized as the value of the initial amplitude $a$ increases. First, we can observe that the value of the critical velocity $v_c$ varies very slowly as the initial amplitude $a$ grows. For instance, $v_c \approx 0.2601$ for $a=0.02$, whereas $v_c\approx 0.2681$ for $a=0.2$, following a linear dependence in $a$ for intermediate values. Second, it can be seen that the 2-bounce windows are deformed as the value of $a$ increases and get broken up in smaller 2-bounce windows. The first 2-bounce window shown in Figure \ref{fig:VelDiaContrafaseFractal} for $a=0.04$ can be used to illustrate this mechanism. This window gets distorted when $a=0.12$ and split into two pieces for $a=0.14$. In turn, one of these pieces is divided again into two new 2-bounce windows for $a=0.20$. Third, spontaneous generation of $n$-bounce windows with $n\geq 2$ can also be identified in the sequence of graphics included in Figure~\ref{fig:VelDiaContrafaseFractal}. For instance, for $a=0.12$ a small 3-bounce window spontaneously emerges in the interval $[0.21585,0.2174]$, which was occupied by the bion formation regime for previous values of $a$. Subsequently, this window is split into two parts, resulting in a 2-bounce window in the middle  for $a=0.14$, which is surrounded by new $n$-bounce windows. This new 2-bounce window gets bigger as $a$ increases and finally splits into two new 2-bounce windows once more, as you can see from the graphics for $a=0.20$. This window generation mechanism could explain the clustering of 2-bounce windows that arise around the  $v_0=0.2566$ value for $a=0.20$.

\begin{figure}[htb]
	\centerline{\includegraphics[height=3.3cm]{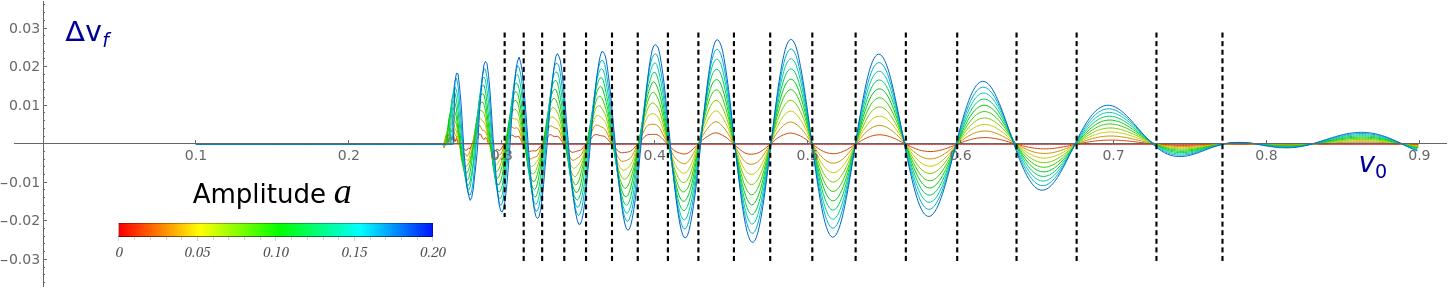}}
	\caption{\small Final velocity difference $\Delta v_f$ of the scattered wobblers as a function of the collision velocity $v_0$ and the initial wobbling amplitude $a$ for the scattering of two wobblers with opposite phase. Recall that $\Delta v_f=0$ for $a=0$ due to spatial reflection symmetry. For the sake of clarity, $n$-bounce processes with $n\geq 2$ have not been included in the plot. The vertical dashed lines mark the zeroes $\widetilde{v}_k$ of the final velocity difference $\Delta v_f$.} \label{fig:VelocityDifferenceContrafase}
\end{figure}

Another important characteristic of this type of scattering processes is that the final velocities of the scattered wobblers are different. This behavior is not surprising because the initial configuration (\ref{configuration04}) is not symmetric. Recall that the initial wobbling phases of the colliding wobblers are different. This velocity difference is very small, and therefore not noticeable in the velocity diagrams shown in Figure~\ref{fig:VelDiaContrafaseAmp020}. In order to emphasize this feature we define the magnitude
\begin{equation}
\Delta v_f = |v_{f,R}|-|v_{f,L}| \, , \label{deltavf}
\end{equation}
as the difference between the final speed $|v_{f,R}|$ of the rightward traveling wobbler and the final speed $|v_{f,L}|$ of the leftward traveling wobbler. Positive values of $\Delta v_f$ imply that the wobbler scattered to the right travels faster than the wobbler scattered to the left, whereas negative values describe the reverse situation. In Figure~\ref{fig:VelocityDifferenceContrafase}, the magnitude $\Delta v_f$ is plotted as a function of the initial velocity $v_0$ and the wobbling amplitude $a$. There, we can see that $\Delta v_f$ has oscillating behavior, which means that there are alternating initial velocity windows in which the wobbler traveling from the left travels faster than the wobbler traveling from the right and vice versa. The amplitudes of the oscillations exhibited by $\Delta v_f$ grow as the value of the parameter $a$ increases. This is reasonable because the vibrational energy stored in the shape mode is greater for bigger values of $a$ and the resonant energy transfer mechanism may deflect a greater amount of this energy to the kinetic energy pool. However, the most remarkable property exhibited by  Figure~\ref{fig:VelocityDifferenceContrafase} is that the zeroes of $\Delta v_f$, the initial velocity values for which the two wobblers disperse with the same velocity, are approximately independent of the initial amplitude $a$. This behavior is precisely followed for sufficiently large values of $v_0$, where the effect of the resonance regime is not noticed (approximately for $v_0\geq 0.3$ in Figure~\ref{fig:VelocityDifferenceContrafase}). 

In Table \ref{ZerosContrafase}, the zeros $\widetilde{v}_k$ of the final velocity difference $\Delta v_f$ (explicitly computed for the case $a=0.04$) are shown in the non-resonance regime. The values $\widetilde{v}_k$ correspond to the nodes of the oscillations found in Figure~\ref{fig:VelocityDifferenceContrafase}, which have been remarked by means of vertical dashed lines.
The location of these points seems to depend mainly on the value of the wobbling phase when the collision between the wobblers occurs. This conjecture is heuristically supported by the following simple argument. Remember that $x_0$ denotes the initial position of the kink center, while $\omega$ represents the wobbling frequency. As previously discussed, the values $x_0=15$ and $\omega=\sqrt{3}$ have been implemented for our numerical simulations. Let $v_0$ be the initial velocity at which the wobblers are initially approaching. In the point particle approximation the collision would happen at the time $t_I=\frac{x_0}{v_0}$. We must bear in mind that there are several factors in the real dynamics which break the precision of this assumption. For example, the interaction between the kinks and/or wobblers can make the collision velocity vary (it is not a constant velocity $v_0$). We shall assume that the phase of the wobbler at the instant $t_I$ can be expressed as
\[
\varphi(v_0)= c(x_0) \, \frac{x_0}{v_0} \omega \sqrt{1-v_0^2} + \delta \, .
\]
where $c(x_0)$ is a correction factor which is included to incorporate the previously mentioned behavior. The main assumption in this case is that $c(x_0)$ does not depend on $v_0$. If we think about the initial impact velocity as a variable $v$, then it makes sense to consider $\varphi(v)=c(x_0) \, \frac{x_0}{v} \omega \sqrt{1-v^2} + \delta$.
Those phenomena depending only on the wobbling phase must exhibit a periodicity based on the relation
\begin{equation}
\varphi(v_0) - \varphi(v)= T\, k\, , \hspace{0.6cm} k\in \mathbb{Z}\,, \label{fase}
\end{equation}
where $T$ is the periodicity associated to our problem. In general, $T=2\pi$ but in the present scenario where we are interested in the zeroes $\widetilde{v}_k$ of $\Delta v_f$ the symmetry of the initial configuration leads to the choice $T=\pi$. From (\ref{fase}) we conclude that the discrete set of velocities
\begin{equation}
f_k(v_0,T)=\frac{v_0 \, x_0 \, \omega}{\sqrt{\frac{k^2 \, T^2\, v_0^2}{4 \,c(x_0)^2}  - \frac{T}{c(x_0)} \, k \, v_0 \, x_0 \, \omega \sqrt{1-v_0^2} + x_0^2 \, \omega^2}} \,,\hspace{0.6cm} k\in \mathbb{Z},
\label{velocities}
\end{equation}
must share similar features. The nodes $\widetilde{v}_k$ of $\Delta v_f$ can be approximately figured out by using equation (\ref{velocities}). In Table \ref{ZerosContrafase} (third column) the values $V_{k}=f_k(v_0,\pi)$, obtained by using the formula (\ref{velocities}) taking as initial input $v_0=\widetilde{v}_0=0.301538$, are included. The value $c(x_0)$ has been adjusted to $c(x_0)= 0.465$. The comparison between the data allows us to conclude that the previous conjecture is satisfied at least for intermediate values of the initial velocity. Of course, the nonlinear nature of the problem makes the argument only an approximation to the actual behavior. This is clear for very large values of the collision velocity. In this regime the amplitudes of the oscillations of $\Delta v_f$ are very attenuated compared to intermediate values of $v_0$. For these cases radiation emission can play a predominant role in the scattering processes.

\begin{table}[htb]
	\centerline{\begin{tabular}{|c|c|c|} \hline
			$k$ & $\widetilde{v}_k$ & $V_k$	\\ \hline
			0 & 0.301538 & 0.301538 \\
			1 & 0.313991 & 0.313224\\
			2 & 0.326057 & 0.325797 \\
			3 & 0.340542 & 0.339354 \\
			4 & 0.354757 & 0.354006 \\
			5 & 0.371738 & 0.369877 \\
			6 & 0.388575 & 0.387110 \\
			\hline
		\end{tabular} \hspace{0.2cm} \begin{tabular}{|c|c|c|} \hline
			$k$ & $\widetilde{v}_k$  & $V_k$	\\ \hline
			7 & 0.408291 & 0.405864 \\ 
			8 & 0.428219  & 0.426323 \\
			9 & 0.451464 & 0.448689 \\
			10 & 0.475173 & 0.473188 \\
			11 & 0.502713 & 0.500069 \\
			12 & 0.531044 & 0.529591 \\
			13 & 0.563882 & 0.562022 \\
			\hline
		\end{tabular} \hspace{0.2cm}
		\begin{tabular}{|c|c|c|} \hline
			$k$ &	$\widetilde{v}_k$  & $V_k$	\\ \hline
			14 & 0.597459 & 0.597606 \\
			15 & 0.636259 & 0.636531 \\
			16 &	0.675558 & 0.678857 \\
			17 &	0.727740 & 0.724418 \\
			18 &	0.770909 & 0.772667 \\ && \\ && \\ \hline
	\end{tabular}}
	\caption{Comparison between the zeros $\widetilde{v}_k$ of the final velocity difference $\Delta v_f$ and the values $V_k=f_k(v_0,\pi)$ obtained by using equation (\ref{velocities}) for the scattering between wobblers with opposite phase and initial wobbling amplitude $a=0.04$. } \label{ZerosContrafase}
\end{table}

At this point it is worthwhile mentioning that the zeroes $\widetilde{v}_k$ introduced in Table  \ref{ZerosContrafase} have been computed when $\delta=0$ in the initial configuration (\ref{configuration04}). The particular location of these points depends on the initial phase $\delta$ introduced in (\ref{configuration04}), although it is clear that the same pattern is periodically reproduced for the values $\delta + k T$ with $k\in \mathbb{Z}$.

\begin{figure}[h]
	\centerline{\includegraphics[height=4.2cm]{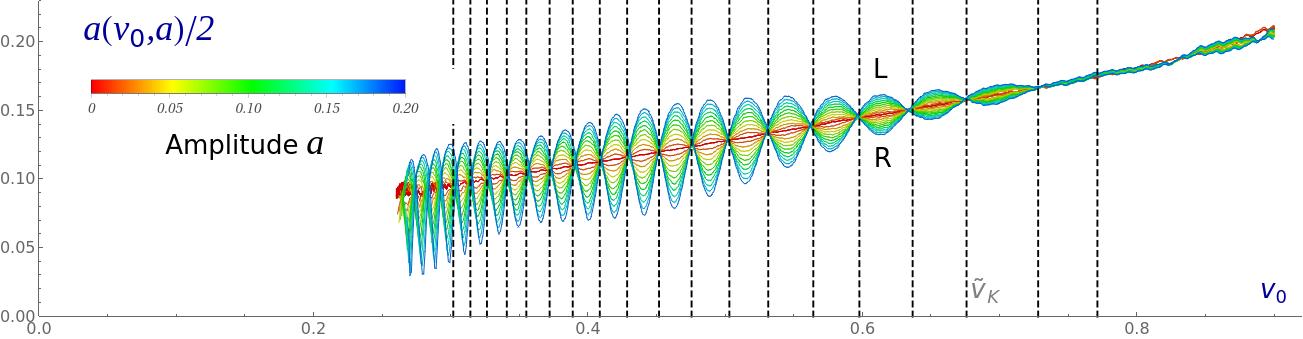}}
	\caption{\small Graphics of the final wobbling amplitudes of the wobblers scattered to the left and to the right as a function of the initial velocity $v_0$ and the initial amplitude $a$. For the sake of clarity, $n$-bounce processes with $n\geq 2$ have not been included. The vertical dashed lines mark the zeroes $\widetilde{v}_k$ of the final velocity difference $\Delta v_f$. The letters $L$ and $R$ label the smooth amplitude functions associated with wobblers traveling left and right, respectively. } \label{fig:AmplitudFinalContrafaseKink1y2}
\end{figure}

Once the final velocities of the scattered wobblers have been examined, we shall now analyze the behavior of the wobbling amplitude of these evolving topological defects. 
In Figure~\ref{fig:AmplitudFinalContrafaseKink1y2}, the oscillation amplitudes of the wobblers moving to the left and to the right are represented as a function of the initial velocity $v_0$ and the initial amplitude $a$. There it can be seen that this magnitude follows an oscillating behavior with respect to the kink-antikink scattering. The variation of these oscillations grows as the parameter $a$ increases. Furthermore, the amplitudes of the resulting wobblers follow an antagonistic behavior. When the oscillation amplitude of the wobbler moving to the left reaches a maximum as a function of the initial velocity $v_0$, the oscillation amplitude of the wobbler moving to the right is minimized and vice versa. The asymmetry of the initial configuration (\ref{configuration04}) causes the wobblers to vibrate at different amplitudes in general. On the other hand, there are some points in the graphs shown in Figure~\ref{fig:AmplitudFinalContrafaseKink1y2} where the amplitudes of the two wobblers coincide. Surprisingly, these points coincide with the zeroes $\widetilde{v}_k$ of the final velocity difference $\Delta v_f$ (as we can observed by means of the vertical dashed lines plotted in Figure~\ref{fig:AmplitudFinalContrafaseKink1y2}). In conclusion, for the initial velocities $\widetilde{v}_k$ the scattered wobblers travel with the same velocity and vibrate with the same wobbling amplitude. 

\begin{figure}[htb]
	\centerline{\includegraphics[height=3.4cm]{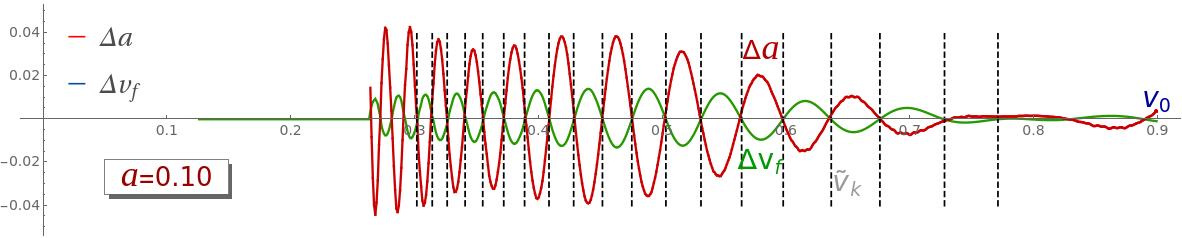}}
	\caption{\small Graphics of $\Delta v_f$ (final velocity difference) and $\Delta a$ (final wobbling amplitude difference) as functions of the initial collision velocity $v_0$ for the scattering between wobblers with opposite phase with $a=0.10$. $n$-bounce processes with $n\geq 2$ have not been included. The vertical dashed lines mark the zeroes $\widetilde{v}_k$ of $\Delta v_f$. } \label{fig:AmplitudYVelocidadContrafase}
\end{figure}

In order to explore the relation between the final velocity and the final wobbling amplitude of the scattered wobblers, we define the amplitude difference 
\begin{equation}
\Delta a = \frac{1}{2} \left[ a_{f,R} -  a_{f,L}  \right]\, ,
\end{equation}
where $a_{f,R}$ and $a_{f,L} $ are, respectively, the final oscillation amplitudes of the wobblers moving to the right and to the left. $\Delta a > 0$ means that the wobbler scattered to the right vibrates strongly than that moving to the left, whereas $\Delta a < 0$ describes the opposite situation. Figure~\ref{fig:AmplitudYVelocidadContrafase} shows simultaneously the final velocity and the amplitude differences $\Delta v_f$ and $\Delta a$, as functions of the initial velocity $v_0$ for the particular value $a=0.10$. It can be seen that when a scattered wobbler gains more kinetic energy than the other, it obtains less vibrational energy, and vice versa. The values $\widetilde{v}_k$ are interpreted as the collision velocities for which the final velocities and the wobbling amplitudes of the scattered wobblers are the same.

\begin{figure}[htb]
	\centerline{\includegraphics[height=3.4cm]{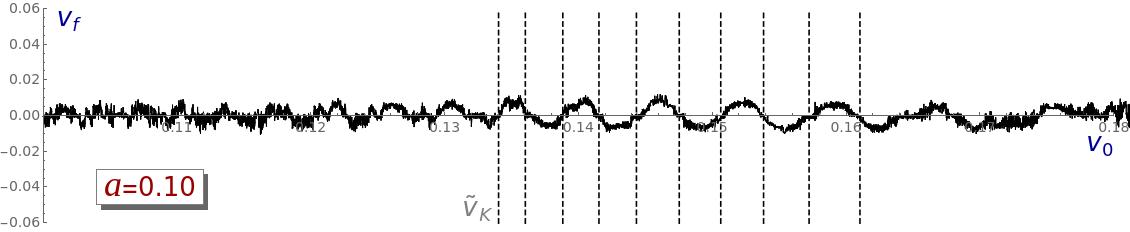}}
	\caption{\small Final bion velocity as a function of the initial velocity $v_0$ in the interval $v_0\in [0.10,0.18]$ for the scattering between two wobblers with opposite phase and the initial wobbling amplitude $a=0.10$. The vertical dashed lines mark some of the nodes of the curve.} \label{fig:VelDiaContrafaseBionAmp010}
\end{figure}

Finally, another consequence of the asymmetry of these scattering events is that the bion (formed as a bound state between the two colliding wobblers) can now move with certain final non-vanishing  velocity after the impact. This velocity will be very small and for this reason it is sometimes difficult to compute its magnitude numerically. In~Figure~\ref{fig:VelDiaContrafaseBionAmp010} the region of the velocity diagram introduced in Figure~\ref{fig:VelDiaContrafaseAmp020} for $a=0.10$ with $v_0\in [0.10,0.18]$ has been enlarged to illustrate the behavior of the bion velocity. 
Again, we find an oscillating pattern, clearly seen in Figure~\ref{fig:VelDiaContrafaseBionAmp010} for the interval $v_0\in [0.13,0.16]$. Also, it turns out that the formula (\ref{velocities}) still governs this oscillating behavior. In the previously mentioned range of $v_0$, vertical dashed lines have been plotted to approximately mark the location of the nodes of the bion velocity. The values used correspond to the initial velocities $v_1 \approx 0.134$, $v_2 \approx 0.1364$, $v_3 \approx 0.1388$, $v_4 \approx 0.1415$, $v_5 \approx 0.1443$, $v_6 \approx 0.1475$, $v_7 \approx 0.1506$, $v_8 \approx 0.1538$, $v_9 \approx 0.1572$ and $v_{10} \approx 0.161$, which can be approximately reproduced by (\ref{velocities}).

\section{Scattering between a kink and a wobbler}
\label{sec:4}

\begin{figure}[htb]
	\centerline{\includegraphics[height=3.5cm]{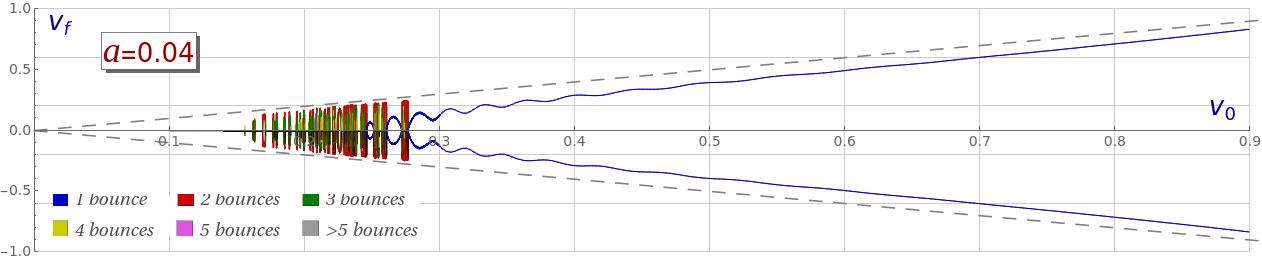}}
\medskip
	\centerline{\includegraphics[height=3.5cm]{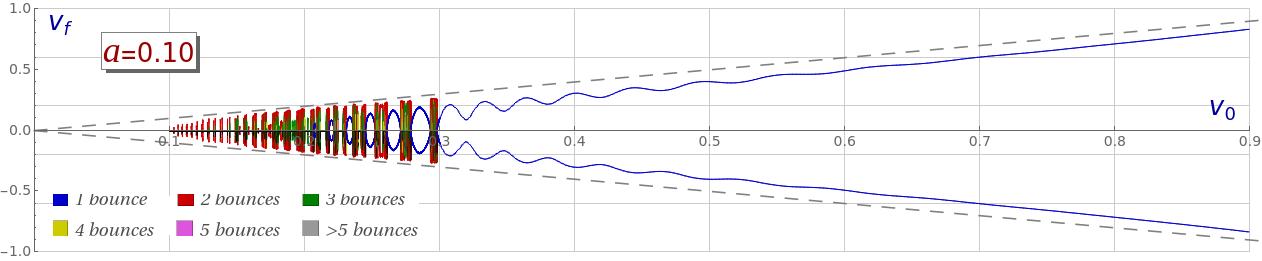}}
\medskip
	\centerline{\includegraphics[height=3.5cm]{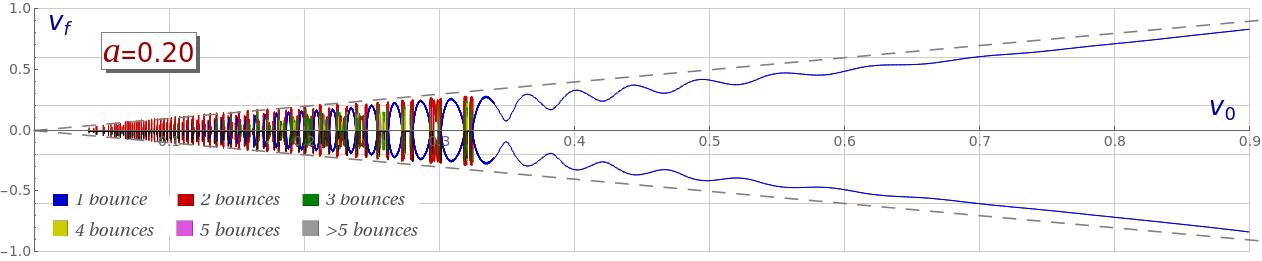}}
	\caption{\small Final versus initial velocity diagram for the wobbler-antikink scattering for the values of the wobbling amplitude $a=0.04$, $a=0.10$ and $a=0.20$. The color code is used to specify the number of bounces suffered by the kinks before escaping.} \label{fig:VelDiaAmp020}
\end{figure}

In this section we shall study the scattering between a wobbler and a kink. This scenario is characterized by the concatenation (\ref{configuration03}). With the first choice of signs, this configuration describes a wobbler and an antikink which travel respectively with velocities $v_0$ and $-v_0$. The rightward traveling wobbler and the leftward traveling antikink approach each other, collide, and bounce back. As usual, the \textit{formation of a bion} and the \textit{reflection of the solutions} complete the list of possible scattering channels. In the reflection regime, the initially unexcited antikink becomes an anti-wobbler after the collision because, in general, the shape mode of this solution is excited. Therefore, after the impact two wobblers emerge moving away with different final velocities in our inertial system. The goal of this study is to analyze the transfer of the vibrational and kinetic energies between the resulting wobblers. The dependence of the final velocities of the scattered extended particles on the initial velocity $v_0$ has been graphically represented in Figure~\ref{fig:VelDiaAmp020} for the cases $a=0.04$, $a=0.1$, and $a=0.2$. 

\begin{figure}[bht]
	\centerline{\includegraphics[height=3cm]{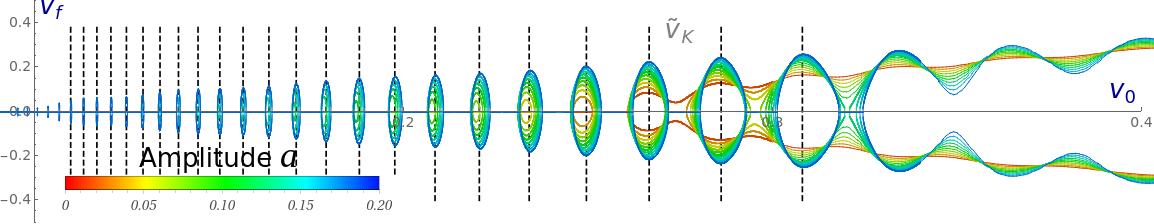}}
	\caption{\small Velocity diagrams for the wobbler-antikink scattering showing the emergence and location of the isolated 1-bounce windows as the value of the wobbling amplitude increases. The vertical dashed lines mark the values of $v_0$ at the center of the 1-bounce windows for the extreme case $a=0.2$. For the sake of clarity, $n$-bounce processes with $n\geq 2$ have not been included. } \label{fig:VelDiaIsolated}
\end{figure}

Some of the most relevant characteristics described in \cite{Alonso2021b} for the scattering between wobbling kinks are also found in this framework, such as the emergence of isolated 1-bounce windows and the growing complexity of the fractal pattern as the initial wobbling amplitude $a$ of the originally rightward-traveling wobbler increases. It is also worthwhile mentioning the presence of oscillations in the 1-bounce tail arising for large values of the initial velocity. However, these features are less accentuated in this scenario. The reason of this behavior lies in the fact that the constructive interference is maximized when the wobblers collide with the same wobbling phase. In particular, we can observe the existence of two isolated 1-bounce windows for the case $a=0.04$. They occupy approximately the region $[0.2458,0.2522] \cup [0.2607,0.2717]$. For $a=0.1$ six of these windows can be identified in $[0.2068,0.2085] \cup [0.2183,0.2214] \cup [0.2310,0.2358] \cup [0.2452,0.2521] \cup [0.2610,0.2709] \cup [0.2787,0.2925]$. Finally, the number of these windows explodes as the initial amplitude $a$ grows. This can be observed in the velocity diagram for $a=0.2$ in Figure \ref{fig:VelDiaAmp020}. Some of the widest 1-bounce windows in this case arise in the set of intervals $[0.2067,0.2108]\cup [0.2185,0.2234] \cup [0.2315,0.2375] \cup [0.2460,0.2535] \cup [0.2621,0.2717] \cup [0.2804,0.2928] \cup [0.3012,0.3179]$. From the previous list of 1-bounce windows, it can be verified that once an isolated 1-bounce window emerges its location is approximately fixed (although its width slightly grows) as the initial wobbling amplitude $a$ increases. This behavior can be checked in Figure \ref{fig:VelDiaIsolated}. Note that the deviation from the rule described above is a small translation of the center of these windows. In Figure \ref{fig:VelDiaIsolated} the vertical dashed lines mark the values of the initial velocity which determine the centers of the 1-bounce windows for the extreme case $a=0.2$. Once again, these velocities approximately follow  relation (\ref{velocities}), which reveals that the role of the phase of the evolving shape mode is predominant in this phenomenon.

\begin{figure}[h]
	\centerline{\includegraphics[height=3.3cm]{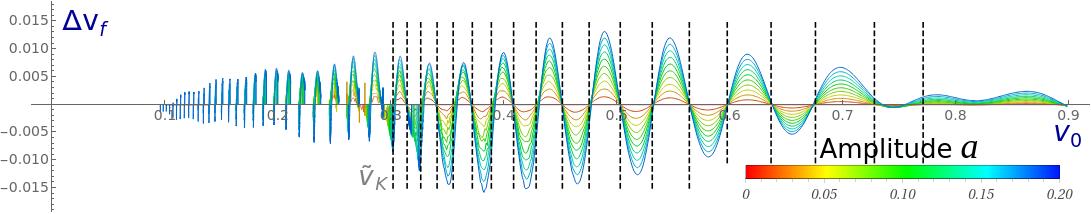}}
	\caption{\small Final velocity difference $\Delta v_f$ of the scattered wobblers as a function of the collision velocity $v_0$ and the initial wobbling amplitude $a$ for the scattering of a wobbler and a kink. $n$-bounce processes with $n\geq 2$ have not been included. The vertical dashed lines mark the zeroes $\widetilde{v}_k$ of $\Delta v_f$ displayed in Table~\ref{ZerosContrafase}. } \label{fig:VelocityDifference}
\end{figure}

The velocity diagrams shown in Figure~\ref{fig:VelDiaAmp020} also have some distinctive properties of their own.
Because the scattering processes introduced in this section are asymmetric, the final velocities of the resulting wobblers are different, as well as their wobbling amplitudes. In order to illustrate this feature more clearly, the difference $\Delta v_f$ between the final speeds of the scattered wobblers is plotted for different values of the wobbling amplitude $a$ in Figure~\ref{fig:VelocityDifference}. For the sake of simplicity, only 1-bounce events have been included in Figure~\ref{fig:VelocityDifference}. As in the case of the scattering between wobblers with opposite phase discussed in Section \ref{sec:3}, the zeros of this function $\Delta v_f$ are approximately independent of the initial amplitude $a$ and, indeed, coincide with the zeroes $\widetilde{v}_k$ introduced in Table~\ref{ZerosContrafase} in Section~\ref{sec:3}. This behavior underlies the fact that the initially rightward wobbler defined in the configuration (\ref{configuration03}) has the same initial conditions as those given by the configuration (\ref{configuration04}). 

\begin{figure}[htb]
\centerline{\includegraphics[height=3.5cm]{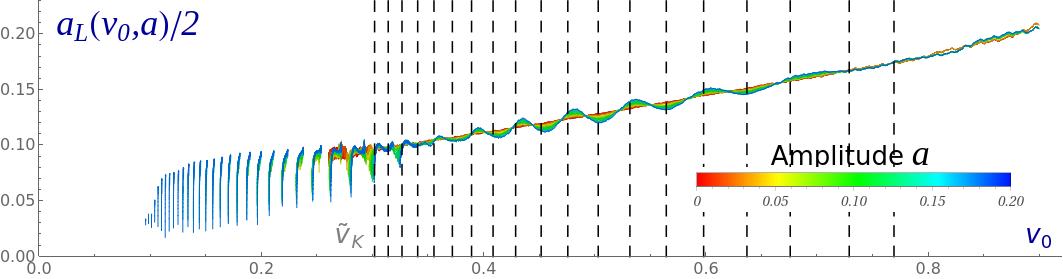}}\medskip
\centerline{\includegraphics[height=3.5cm]{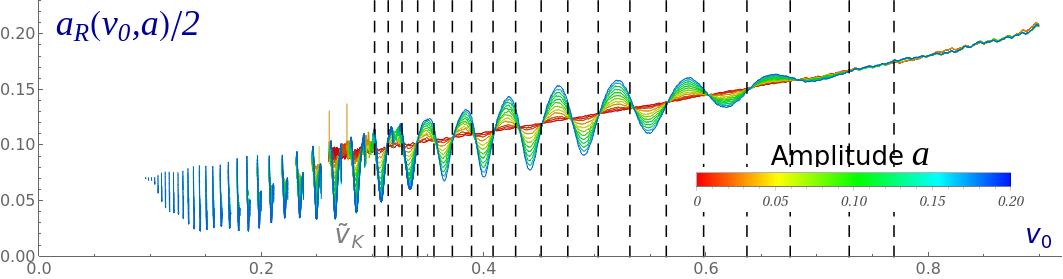}}
	\caption{\small Final wobbling amplitudes $a_L$ and $a_R$ of the wobblers scattered to the left (top) and to the right (bottom) as a function of the initial velocity $v_0$ and the initial wobbling amplitude $a$ of the colliding wobbler. The vertical dashed lines mark the zeroes $\widetilde{v}_k$ of $\Delta v_f$ displayed in Table \ref{ZerosContrafase}.} \label{fig:AmplitudFinalTotalKink}
\end{figure}

In Figure~\ref{fig:AmplitudFinalTotalKink} the final wobbling amplitudes of the scattered wobblers are plotted as a function of the initial velocity $v_0$ and the initial wobbling amplitude $a$. Recall that $a_L(v_0,a)$ and $a_R(v_0,a)$ represent, respectively, the final wobbling amplitudes of the resulting leftward and rightward traveling wobblers after the collision. We can observed that the shape modes of the scattered wobblers become excited and its amplitudes are similar as a function of the initial velocity, oscillating around the values found for the kink-antikink scattering events (with $a=0$). However, the amplitude of these oscillations is much bigger for the final rightward traveling wobbler.

\begin{figure}[h]
	\centerline{\includegraphics[height=3.5cm]{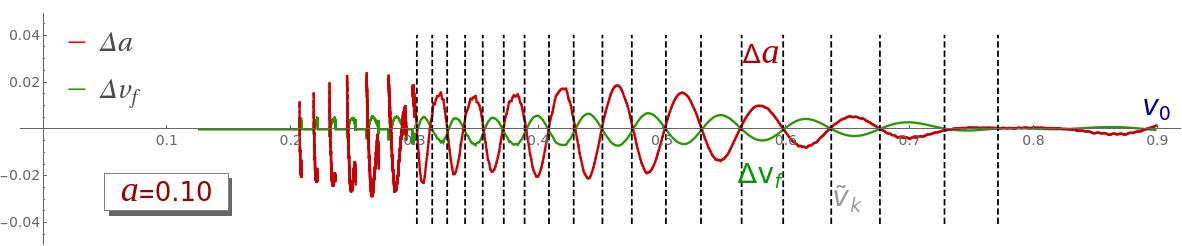}}
	\caption{\small Graphics of $\Delta v_f$ (final velocity difference) and $\Delta a$ (final wobbling amplitude difference) as a function of the initial collision velocity $v_0$ for the scattering between a wobbler and an antikink with $a=0.10$. $n$-bounce processes with $n\geq 2$ have not been included. The vertical dashed lines mark the zeroes $\widetilde{v}_k$ of $\Delta v_f$.} \label{fig:AmplitudFinal020Kink1y2}
\end{figure}

To illustrate the role of the the zeroes $\widetilde{v}_k$ of the final velocity difference $\Delta v_f$ shown in Table~\ref{ZerosContrafase} in this scenario, the functions $\Delta v_f$ and $\Delta a$ have been represented simultaneously for the case $a=0.10$ in Figure~\ref{fig:AmplitudFinal020Kink1y2}. As in the scattering between wobblers with opposite phase, the values $\widetilde{v}_k$ determine the initial velocities for which the final velocities and the final wobbling amplitudes are the same for the both scattered wobblers.

\section{Conclusions}
\label{sec:5}

This paper delves into the study on the scattering between wobbling kinks initially addressed in \cite{Alonso2021b}. Here, we have investigated the asymmetric scattering between kinks and wobblers (kinks whose shape mode is excited) in the standard $\phi^4$ model. In particular, two different scenarios in this context have been considered: (a) the scattering between wobblers with opposite phases, and (b) the scattering between a wobbler and an unexcited antikink. Both cases exhibit the usual bion formation and reflection regimes, which are infinitely interlaced forming a fractal structure embedded in the final versus initial velocity diagram. However, the first case involves a destructive interference of the shape modes in the collision. As a consequence, the growth in the complexity of the fractal pattern is smaller than that found in \cite{Alonso2021b}, where the colliding wobbling kinks travel with the same phase leading to a constructive interference at the impact. For example, the emergence of isolated 1-bounce windows is not found in this new case (at least for moderate values of the initial wobbling amplitude $a$), although the splitting of $n$-bounce widows is present. On the other hand, the kink scattering in the second scenario displays similar features (although more attenuated) than to those found in \cite{Alonso2021b}. 

Due to the asymmetry of the initial configurations (\ref{configuration04}) and (\ref{configuration03}), the final velocities and wobbling amplitudes of the scattered wobblers are different in general. However, there is a sequence of initial velocities for which both the final velocities and wobbling amplitudes coincide. These values are almost independent of the initial wobbling amplitude $a$ when the initial wobbling phase considered in (\ref{configuration04}) and (\ref{configuration03}) is fixed. Besides, the values of these velocities very approximately follow the expression (\ref{velocities}). This means that the phase associated to the shape modes of the evolving wobblers at the collision instant plays a predominant role in the scattering properties of these objects. 
Indeed, (\ref{velocities}) allows to obtain values of the initial velocities which share similar features. For example, this expression has been used in the second scenario to predict the location of the maxima of the isolated 1-bounce windows. Finally, it is also worthwhile mentioning the results displayed in Figures \ref{fig:AmplitudYVelocidadContrafase} and \ref{fig:AmplitudFinal020Kink1y2}. It can be verified that systematically when a scattered wobbler gains more kinetic energy than the other, it obtains less vibrational energy and vice versa.

The research introduced in the present work opens up some possibilities for future work. For example, the $\phi^6$ model implies a  resonance regime similar to the $\phi^4$ model, although it does not present vibrational eigenstates in the second-order small fluctuation operator. The characteristics of scattered wobbling kinks can be analyzed to study their influence on the resonant energy transfer mechanism. Alternatively, you can build a  model twin to the $\phi^6$ model that involves internal modes. By doing this, we could compare the scattering processes of the twin model with those of the standard $\phi^6$ model. In this way, it will be possible to examine the role that shape modes play in the collision process.
Furthermore, many other different topological defects (kinks in the double sine-Gordon model, deformed $\phi^4$ models, hybrid and hyperbolic models, etc.) could be studied in the new perspective presented here. Work in these directions is in progress.

\section*{Acknowledgments}

A. Alonso-Izquierdo acknowledges Spanish MCIN financial support under grant PID2020-113406GB-I0. He also acknowledges the Junta de Castilla y Le\'on for financial support under grants SA067G19. L.M. Nieto acknowledges Spanish MCIN financial support under grant PID2020-113406GB-I0. This research has made use of the high performance computing resources of the Castilla y Le\'on Supercomputing Center (SCAYLE, www.scayle.es), financed by the European Regional Development Fund (ERDF).


\begin{thebibliography}{99}

\addcontentsline{toc}{section}{References}

\bibitem{Manton2004} 
N. Manton, \textit{Topological Solitons} (Cambridge University Press, Cambridge,  2004).

\bibitem{Shnir2018} 
Y. M. Shnir, \textit{Topological and non-topological solitons in scalar field theories} (Cambridge University Press, Cambridge, 2018).

\bibitem{Kevrekidis2019} 
P. G. Kevrekidis and J. Cuevas-Maraver (eds), {\it A Dynamical Perspective on the $\phi^4$ Model. Nonlinear Systems and Complexity} Vol.~26 (Springer, Cham, 2019). 

\bibitem{Sugiyama1979} 
T. Sugiyama, 
Progr. Theoret. Phys. \textbf{61},  1550 (1979).

\bibitem{Campbell1983} 
D. K. Campbell, J. S. Schonfeld and C. A. Wingate, 
Phys. D \textbf{9}, 1 (1983).

\bibitem{Anninos1991} 
P.  Anninos, S. Oliveira and R. A. Matzner, 
Phys. Rev. D \textbf{44}, 1147 (1991).

\bibitem{Eschenfelder1981} 
A. H. Eschenfelder, {\it Magnetic Bubble Technology} (Springer-Verlag, Berlin, 1981).

\bibitem{Jona1993} 
F. Jona and G. Shirane, {\it Ferroelectric Crystals} (Dover, New York, 1993).

\bibitem{Strukov} 
B. A. Strukov and A. P. Levanyuk, {\it Ferroelectric Phenomena in Crystals: Physical Foundations} (Springer-Verlag, Berlin, 1998).

\bibitem{Vilenkin1994} 
A. Vilenkin and E. P. S. Shellard, \textit{Cosmic strings and other topological defects} (Cambridge University Press, Cambridge, 1994).

\bibitem{Vachaspati2006} 
T. Vachaspati, \textit{Kinks and Domain walls: An Introduction to classical and quantum solitons} (Cambridge University Press, Cambridge, 2006).

\bibitem{Mollenauer2006} 
L. F. Mollenauer and J. P. Gordon, \textit{Solitons in optical fibers--Fundamentals and applications} (Academic Press, Burlington, 2006).

\bibitem{Schneider2004} 
T. Schneider, \textit{Nonlinear optics in Telecommunications} (Springer, Heidelberg, 2004).

\bibitem{Agrawall1995} 
G. P. Agrawall, \textit{Nonlinear Fiber Optics} (Academic Press, San Diego, 1995).

\bibitem{Davydov1985} 
A. S. Davydov, \textit{Solitons in molecular systems} (D. Reidel, Dordrech, 1985).

\bibitem{Bazeia1999} 
D. Bazeia and E. Ventura, 
Chem. Phys. Lett. \textbf{303}, 341 (1999).

\bibitem{Yakushevich2004} 
L. V. Yakushevich, \textit{Nonlinear Physics of DNA} (Wiley-VCH, Weinheim, 2004).

\bibitem{Shiefman1979} 
J. Shiefman and P. Kumar, 
Phys. Scr. \textbf{20}, 435 (1979).

\bibitem{Peyrard1983} 
M. Peyrard and D. K. Campbell, 
Physica D \textbf{9}, 33 (1983).

\bibitem{Goodman2005} 
R. H. Goodman and R. Haberman, 
SIAM J. Appl. Dyn. Syst. \textbf{4}, 1195 (2005).

\bibitem{Gani1999} 
V. A. Gani and A. E. Kudryavtsev, 
Phys. Rev. E \textbf{60}, 3305 (1999).

\bibitem{Malomed1989} 
B. A. Malomed, 
Phys. Lett. A \textbf{136}, 395 (1989).

\bibitem{Gani2018} 
V. A. Gani, A. M. Marjaneh, A. Askari, E. Belendryasova, and D. Saadatmand, 
Eur. Phys. J. C \textbf{78}, 345 (2018).

\bibitem{Gani2019} 
V. A. Gani, A. M. Marjaneh and D. Saadatmand, 
Eur. Phys. J. C. \textbf{79}, 620 (2019). 

\bibitem{Simas2016} 
F. C. Simas, A .R. Gomes, K.Z . Nobrega, and J.C.R.E. Oliveira, 
J. High Energy Phys.  {9} (2016) 104.

\bibitem{Gomes2018} 
A. R. Gomes, F. C. Simas, K. Z. Nobrega, and P. P. Avelino, 
J. High Energy Phys. {10} (2018) 192.

\bibitem{Bazeia2017b} 
D. Bazeia, E. Belendryasova and V. A. Gani, 
J. Phys.: Conf. Ser. \textbf{934}, 012032 (2017).

\bibitem{Bazeia2017a} 
D. Bazeia, E. Belendryasova and V. A. Gani, 
 Eur. Phys. J. C \textbf{78}, 340 (2018).

\bibitem{Bazeia2019} 
D. Bazeia, A. R. Gomes, K. Z. Nobrega, and F. C. Simas, 
Int. J. Mod. Phys. A \textbf{34}, 1950200 (2019). 

\bibitem{Adam2019} 
C. Adam, K. Oles, T. Romanczukiewicz, and A. Wereszczynski, 
Phys. Rev. Lett. \textbf{122}, 241601 (2019).

\bibitem{Romanczukiewicz2018} 
T. Romanczukiewicz and Y. Shnir, \textit{Some recent developments on kink collisions and related topic: A Dynamical Perspective on the $\phi^4$ Model} (Springer, Cham, 2019).

\bibitem{Adam2020} 
C. Adam, K. Oles, T. Romanczukiewicz, and A. Wereszczynski, 
Phys. Rev. D \textbf{101}, 105021 (2020). 

\bibitem{Mohammadi2020} 
M. Mohammadi and R. Dehghani, 
Commun. Nonlinear Sci. Numer. Simulat. {\bf 94}, 105575 (2021).

\bibitem{Yan2020} 
H. Yan, Y. Zhong, Y. X. Liu, and K. Maeda, 
Phys. Lett. B \textbf{807}, 135542 (2020).

\bibitem{Romanczukiewicz2017} 
T. Romanczukiewicz, 
Phys. Lett. B \textbf{773}, 295 (2017).

\bibitem{Weigel2014} 
H. Weigel, 
J. Phys.: Conf. Ser. \textbf{482}, 012045 (2014).

\bibitem{Gani2014} 
V. A. Gani, A. E. Kudryavtsev and M. A. Lizunova, 
Phys. Rev. D \textbf{89}, 125009 (2014).

\bibitem{Bazeia2018b} 
D. Bazeia, A. R. Gomes, K. Z. Nobrega, and F.C. Simas, 
Phys. Lett. B \textbf{793}, 26 (2019).

\bibitem{Lima2019} F.C. Lima, F.C. Simas, K.Z. Nobrega and A.R. Gomes, 
J. High Energy Phys. {10} (2019) 147.

\bibitem{Marjaheh2017} 
A. M. Marjaneh, V. A. Gani, D. Saadatmand, S. V. Dmitriev, and K. Javidan, 
J. High Energy Phys.  {07} (2017) 028.

\bibitem{Belendryasova2019} 
E. Belendryasova and V. A. Gani, 
Commun. Nonlinear Sci. Numer. Simulat. \textbf{67}, 414 (2019). 

\bibitem{Zhong2020} 
Y. Zhong, X. L. Du, Z. C. Jiang, Y. X, Liu and Y. Q. Wang, 
J. High Energy Phys.  {02} (2020) 153.

\bibitem{Bazeia2020c} 
D. Bazeia, A. R. Gomes, K. Z. Nobrega and F. C. Simas, 
Phys. Lett. B \textbf{803}, 135291 (2020).

\bibitem{Christov2019} 
I. C. Christov, R. J. Decker, A. Demirkaya, V. A. Gani, P. G. Kevrekidis, and R. V. Radomskiy, 
Phys. Rev. D \textbf{99}, 016010 (2019).

\bibitem{Christov2019b} 
I. C. Christov, R. J. Decker, A. Demirkaya, V. A. Gani, P. G. Kevrekidis, A. Khare, and A. Saxena, 
Phys. Rev. Lett. \textbf{122}, 171601 (2019).

\bibitem{Christov2020} 
I. C. Christov, R. J. Decker, A. Demirkaya, V. A. Gani, P. G. Kevrekidis, and A. Saxena, 
Commun. Nonlinear Sci. Numer. Simulat. \textbf{97}, 105748 (2021).

\bibitem{Halavanau2012} 
A. Halavanau, T. Romanczukiewicz and Ya. Shnir, 
Phys. Rev. D \textbf{86}, 085027 (2012).

\bibitem{Romanczukiewicz2008} 
T. Romanczukiewicz, 
Acta Phys. Polon. B \textbf{39}, 3449 (2008).

\bibitem{Alonso2018} 
A. Alonso-Izquierdo, 
Phys. Rev. D \textbf{97}, 045016 (2018).

\bibitem{Alonso2018b} 
A. Alonso-Izquierdo, 
Phys. Scr. \textbf{94}, 085302 (2019).

\bibitem{Alonso2017} 
A. Alonso-Izquierdo, 
Physica D: Nonlinear Phenomena \textbf{365}, 12 (2017).

\bibitem{Alonso2019} 
A. Alonso-Izquierdo, 
Commun. Nonlinear Sci. Numer. Simulat. \textbf{75}, 200 (2019).

\bibitem{Alonso2020} 
A. Alonso-Izquierdo, 
Commun. Nonlinear Sci. Numer. Simulat. \textbf{85}, 105251 (2020).

\bibitem{Alonso2021} 
A. Alonso-Izquierdo, M.A. Gonzalez Leon, J. Martin Vaquero, M. de la Torre Mayado, 
Commun. Nonlinear Sci. Numer. Simulat. \textbf{103}, 106011 (2021).

\bibitem{Ferreira2019} 
L. A. Ferreira, P. Klimas and W. J. Zakreswski, 
J. High Energy Phys.  {01} (2019) 020.

\bibitem{Goodman2002} 
R. H. Goodman, P. J. Holmes and M. I. Weinstein, 
Physica D \textbf{161}, 21 (2002).

\bibitem{Goodman2004} 
R. H. Goodman and R. Haberman, 
Phys. D \textbf{195}, 303 (2004).

\bibitem{Malomed1985}  
B. A. Malomed, 
Physica D: Nonlinear Phenomena \textbf{15}, 385 (1985).

\bibitem{Malomed1992} 
B. A. Malomed, 
J. Phys. A: Math. Gen. \textbf{25}, 755 (1992).

\bibitem{Saadatmand2015} 
D. Saadatmand, S.  V. Dmitriev, D. I. Borisov, P. G. Kevrekidis, M. A. Fatykhov, and K. Javidan, 
Commun. Nonlinear Sci. Numer. Simulat. \textbf{29}, 267 (2015).

\bibitem{Saadatmand2018} 
D. Saadatmand, D. I. Borisov, P. G. Kevrekidis, K. Zhou, and S.V. Dmitriev, 
Commun. Nonlinear Sci. Numer. Simulat. \textbf{56}, 62 (2018).

\bibitem{Manton1997} 
N. S. Manton and H. Merabet, 
Nonlinearity \textbf{10}, 3 (1997).

\bibitem{Adam2018} 
C. Adam, T. Romanczukiewicz and A. Wereszczynski, 
J. High Energy Phys.  {3} (2019) 131.

\bibitem{Adam2019b} 
C. Adam, K. Oles, J. M. Queiruga, T. Romanczukiewicz, and A. Wereszczynski, 
J. High Energy Phys.  {07} (2019) 150.

\bibitem{Adam2020b} 
C. Adam, K. Oles, T. Romanczukiewicz and A. Wereszczynski, 
Phys. Rev. E \textbf{102} (2020) 062214.

\bibitem{Dorey2011} 
P. Dorey, K. Mersh, T. Romanczukiewicz and Y. Shnir, 
Phys. Rev. Lett. \textbf{107}, 091602 (2011).

\bibitem{Dorey2018} 
P. Dorey and T. Romanczukiewicz, 
Phys. Lett. B \textbf{779}, 117 (2018).


\bibitem{Mohammadi2021b} 
M. Mohammadi, R. Dehghani. 
Commun. Nonlinear Sci. Numer. Simulat. \textbf{94}, 105575 (2021).


\bibitem{Campos2020} 
J. G. F. Campos and A. Mohammadi, 
Eur. Phys. J. C, \textbf{80}, 352 (2020).

\bibitem{Blanco-Pillado2021} 
J.J. Blanco-Pillado,  D. Jimenez-Aguilar, J. Urrestilla,
JCAP01 (2021) 027.


\bibitem{Takyi2016} 
I. Takyi and H. Weigel, 
Phys. Rev. D \textbf{94}, 085008 (2016).

\bibitem{Pereira2020} 
C. F. S. Pereira, G. Luchini, T. Tassis, and C.P. Constantinidis, 
J. Phys. A: Math. Theor. {\bf 54}, 075701 (2021).

\bibitem{Manton2021} 
N. S. Manton, K. Oles, T. Romanczukiewicz, and A. Wereszczynski, 
Phys. Rev. Lett. \textbf{127}, 071601 (2021).

\bibitem{Barashenkov2009} 
I. V. Barashenkov and O. F. Oxtoby, 
Phys. Rev. E \textbf{80}, 026608 (2009).

\bibitem{Barashenkov2018} 
I. V. Barashenkov, in \textit{The Continuing Story of the Wobbling Kink}, edited by P. Kevrekidis and J. Cuevas-Maraver, {\it A Dynamical Perspective on the $\phi^4$ Model. Nonlinear Systems and Complexity} Vol.~26 (Springer, Cham, 2019). 

\bibitem{Segur1983} 
H. Segur, 
J. Math. Phys. \textbf{24}, 1439 (1983).

\bibitem{Alonso2021b} 
A. Alonso-Izquierdo, J. Queiroga-Nunes, L.M. Nieto, 
Phys. Rev. D \textbf{103}, 045003 (2021).

\bibitem{Campos2021} 
J. G. F. Campos and A. Mohammadi, 
\textit{Wobbling double sine-Gordon kinks}, arXiv. 2103.04908 (2021).

\end{thebibliography}
\end{document}